\documentclass[twocolumn,aps,prb,superscriptaddress,nofootinbib]{revtex4-1}
\usepackage{graphicx}
\usepackage{times}
\usepackage{bm}
\usepackage[linktocpage=true,colorlinks=true,urlcolor=blue]{hyperref}
\usepackage{pgfplots}
\usepackage{amsfonts}
\usepackage{amsmath}
\usepackage{xcolor}
\usepackage{braket}
\usepackage{relsize}
\usetikzlibrary{arrows}
\usepackage{enumerate}
\usetikzlibrary[patterns] 
\usetikzlibrary[snakes] 
\usetikzlibrary[shapes] 
\usepackage{pgfplots}
\usepackage{newfloat}
\DeclareFloatingEnvironment[name={Supplementary Figure}]{suppfigure}

\graphicspath{ {Figures/} }

\begin{document}

\title{Subband occupation in semiconductor-superconductor nanowires}

\author{Benjamin D. Woods}
\affiliation{Department of Physics and Astronomy, West Virginia University, Morgantown, WV 26506, USA}
\author{Sankar Das Sarma}
\affiliation{Condensed Matter Theory Center and Joint Quantum Institute, Department of Physics, University of Maryland, College Park, Maryland, 20742-4111, USA}
\author{Tudor D. Stanescu}
\affiliation{Department of Physics and Astronomy, West Virginia University, Morgantown, WV 26506, USA}
\affiliation{Condensed Matter Theory Center and Joint Quantum Institute, Department of Physics, University of Maryland, College Park, Maryland, 20742-4111, USA}

\begin{abstract}
Subband occupancy (i.e. the number of occupied subbands or energy levels in the semiconductor) is a key physical parameter characterizing the topological properties of superconductor-semiconductor hybrid systems in the context of the search for non-Abelian Majorana zero modes. We theoretically study the subband occupation of semiconductor nanowire devices as function of the applied gate potential, the semiconductor-superconductor (SM-SC) work function difference, and the surface charge density by solving self-consistently the Schr\"{o}dinger-Poisson equations for the conduction electrons of the semiconductor nanowire. Realistic surface charge densities, which are responsible for band bending, are shown to significantly increase the number of occupied subbands, making it difficult or impossible to reach a regime where only a few subbands are occupied. We also show that the energy separation between subbands is significantly reduced in the regime of many occupied subbands, with highly detrimental consequences for the realization and observation of robust Majorana zero modes. As a consequence, the requirements for the realization of robust topological superconductivity and Majorana zero modes should include a low value of the chemical potential, consistent with the occupation of only a few subbands. Finally, we show that the local density of states on the exposed nanowire facets provides a powerful tool for identifying a regime with many occupied subbands and is capable of providing additional critical information regarding the feasibility of Majorana physics in semiconductor-superconductor devices. In our work, we address both InAs/Al and InSb/Al superconductor-nanowire hybrid systems of current experimental interest.
\end{abstract}

\maketitle

\section{Introduction}
Majorana zero modes (MZMs) localized at the edges of topological superconductors serve as a promising platform for topological quantum computation and simulation \cite{Bravyi2002,Nayak2008,Lahtinen2017,Wille2019,You2019,Alicea2012,Beenakker2013,Lutchyn2018}. Following initial theoretical proposals \cite{Sau2010a,Oreg2010,Alicea2010,Sau2010,Lutchyn2010}, there has been significant experimental progress towards their realization and manipulations within semiconductor-superconductor hybrid nanostructures \cite{Mourik2012,Deng2012,Das2012,Albrecht2016,Chen2017,Suominen2017,Zhang2018,Grivnin2018,Vaitiekenas2018,Krizek2018}. While many experimental results are consistent with the presence of Majorana zero modes, there are concerns that at least some of the observations of zero-bias conductance peaks are actually associated with non-topological low-energy Andreev bound states (ABSs). Such low-energy ABSs can be produced by  several sources, including soft confinement \cite{Kells2012,Stanescu2018b}, inhomogeneous superconducting pairing \cite{Fleckenstein2018}, disorder \cite{Bagrets2012,Liu2012,DeGottardi2013,Adagideli2014}, accidental quantum dots \cite{Liu2017a,Lai2019}, and inter-subband coupling \cite{Woods2019b}. This is especially concerning in the case of trivial low-energy states that mimic very faithfully the local phenomenology of Majorana zero modes, the so-called quasi-Majorana \cite{Vuik2019}, or partially-separated ABS states \cite{Moore2018}. 

While some of the ABS-producing mechanisms such as, for example, long-range inhomogeneous potentials (e.g., soft confinement) and inhomogeneous paring, can be understood within a strictly one dimensional model, other mechanisms rely on or are strongly enhanced by the presence of multiple occupied subbands in wires with finite thickness. For example, the inter-band coupling mechanism produces low-energy ABSs  though mixing multiple subbands in the presence of, e.g., an inhomogeneous electrostatic potential \cite{Woods2019b,Chen2019}. Note that this is a specific example of trivial low-energy states that emerge generically in systems with only particle-hole symmetry \cite{Pan2019} (i.e. systems in the D symmetry class). Also note that, at high occupancy, the presence  of an inhomogeneous effective potential (generated by electrostatic gates, stain, etc.) is similar to the presence of disorder.
In all cases, the subband mixing is expected to be strong when many subbands are occupied and the energy spacing between subbands is small. Consequently, the many-subband regime is characterized by a proliferation of topologically-trivial low-energy states and detrimental to the realization of robust MZMs. 
In particular, for high subband occupancy, the Hamiltonian describing the semiconductor-superconductor hybrid system is better represented by a class D random matrix even without any explicit disorder, since the inter-subband couplings are likely to change drastically because of small variations in the spin-orbit coupling and background potential in the system.  Such a random matrix for high subband occupancy then will generically produce trivial zero bias conductance peaks mimicking Majorana zero modes with these trivial zero bias peaks having no topological significance at all \cite{Pan2019}. Thus, high subband occupancy by itself works as an effective emergent disorder for the topological properties of the system even in the absence of explicit random disorder (similar to how nuclear energy levels for many nucleon nuclei can be represented by random matrices even though there is no explicit randomness in the problem \cite{Wigner1955}).

There are thus many reasons to avoid high occupancy in semiconductor nanowires while searching for topological Majorana zero modes.  It is therefore important to theoretically establish the criteria for subband occupancy in the experimentally relevant nanowires.. To establish whether or not a hybrid device is in the many-subband regime, it is essential to determine the expected number of occupied subbands and the corresponding characteristic subband spacing. In this work we address this problem by studying the subband occupation of InAs/Al nanowire devices as function of the relevant system parameters (see appendix \ref{InSb} for InSb/Al hybrid systems). More specifically, we solve self-consistently the Schr\"{o}dinger-Poisson equations associated with the conduction electrons for a semiconductor-superconductor device with a geometry qualitatively similar to that of hybrid devices used in current experiments. 
A critical feature that we include in our model is a finite surface charge density on the InAs nanowire, which is known to have an accumulation of surface charge, apparently due to surface point defects \cite{Olsson1996}. While several studies have investigated devices using a Schr\"{o}dinger-Poisson framework \cite{Vuik2016,Woods2018,Antipov2018,Mikkelsen2018,Winkler2019,Escribano2019}, only some have included the effects of surface charge \cite{Winkler2019,Escribano2019}. Here we focus on understanding the possible effect of the surface charge on the subband occupation and the inter-subband energy spacing. We find that regimes characterized by many occupied subbands are possible and even likely within realistic windows of system parameters. On the other hand, gating the system to reach the few-subbands regime is usually not possible (except for rather narrow parameter windows) due to the onset of holes. Furthermore, the many-subbands regime is typically associated with small values of the inter-subband spacing near the Fermi level, which has highly detrimental consequences for the realization of robust MZMs. Finally, to assess whether a device is in the optimal few-subbands or the undesirable many-subband regimes, 
we propose a measurement of the LDOS on the exposed facets of the nanowire, e.g., using an STM. We calculate the expected characteristic zero-energy LDOS features obtained by varying the applied gate potential within different regimes of device parameters and asses the capabilities and limitations of such a measurement.

The remainder of this paper is organized as follows. In Sec. II we discuss the hybrid semiconductor-superconductor device, the model used to described its electronic properties, and the methods used in the calculations. The results of our numerical analysis are presented in Sec. III. These results include  the dependence of the subband occupation,  average subband separation near the Fermi level, and local density of states on relevant system parameters. Our conclusions are presented in Sec. IV. The three appendices provide theoretical details on the Luttinger model for holes (App. \ref{Lut}), the InSb nanowires (App. \ref{InSb}), and the local density of states (App. \ref{LDOS}).

\begin{figure}[t]
\begin{center}
\includegraphics[width=0.4\textwidth]{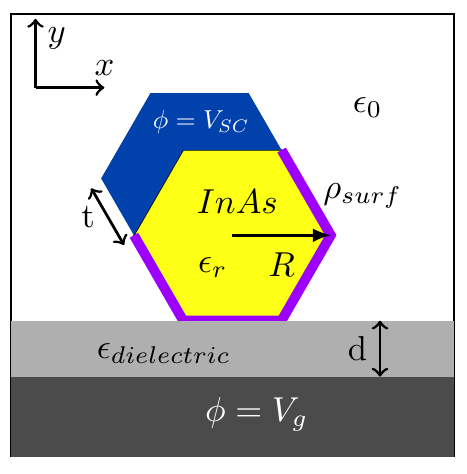}
\end{center}
\vspace{-7mm}
\caption{Transverse profile of the hybrid device studied in this work. An InAs nanowire (yellow) of radius R is partially covered by a superconductor (blue) and placed on an insulating substrate (light gray). The device is gated from below (dark gray) to tune the electrostatic potential. A uniform surface charge (purple) is placed on the InAs surfaces other than the SM-SC interface.}
\label{FIG1}
\vspace{-1mm}
\end{figure}

\section{Device and modeling}

In this work we focus on quasi one-dimensional (1D) semiconductor-superconductor (SM-SC) devices having a transverse cross-section as shown schematically in Fig. \ref{FIG1}. For concreteness, we consider zinc-blende-structured InAs semiconductor wires partially covered by a superconductor, e.g., Al, grown epitaxially on two faces of the nanowire. The proximitized wire is gated from below, to be able to tune the electrostatic potential within the system, and considered  to be infinitely long. To estimate the dependence of the occupancy of confinement-induced conduction subbands on the applied gate potential, we model the nanowire using a simple effective mass model given by the Hamiltonian 
\begin{equation}
H = \frac{\hbar^2}{2 m^{*}}\left(-\partial_x^2 - \partial_y^2 + k_z^2\right) - \mu - e\phi\left(x,y\right), \label{Ham}
\end{equation}
where $m^*$ is the effective mass, $\mu$ is the chemical potential, $e$ is the elementary charge, and $\phi$ is the electrostatic potential inside the wire. We {stress} that, while spin-orbit coupling, Zeeman splitting, and proximity-induced superconductivity are essential for the emergence of  Majorana zero modes at the ends of the wire, they are practically irrelevant {on the energy scale that determines the subband occupancy}. Therefore, for simplicity (and to keep the number of non-essential unknown parameters at a minimum),  we neglect these contributions to the Hamiltonian. The electrostatic potential $\phi\left(x,y\right)$ in Eq.(\ref{Ham}) is determined by the Poisson equation
\begin{equation}
\nabla \cdot \epsilon\nabla \phi \left(\vec{r}\right) = \rho\left(\vec{r}\right), \label{Pois}
\end{equation}
where $\epsilon$ is the material dependent dielectric constant, and $\rho$ is the charge density of the system. This charge density includes contributions from the conduction electrons, as well as possible surface charges \cite{Olsson1996}, and has the generic form 
\begin{equation}
\rho(\vec{r}) = \rho_{f}(\vec{r}) + \rho_{surf}(\vec{r}),
\end{equation}
where $\rho_{f}$, which is given by the occupied eigenstates of Eq.(\ref{Ham}), has to be calculated self-consistently, while $\rho_{surf}$ represents an immobile surface charge density. Following Ref. \cite{Winkler2019}, we model this surface charge as a uniform layer of (positive) charge of thickness $\ell$ on the nanowire surfaces other than the semiconductor-superconductor interface (see Fig. \ref{FIG1}). 
\begin{figure}[t]
\begin{center}
\includegraphics[width=0.48\textwidth]{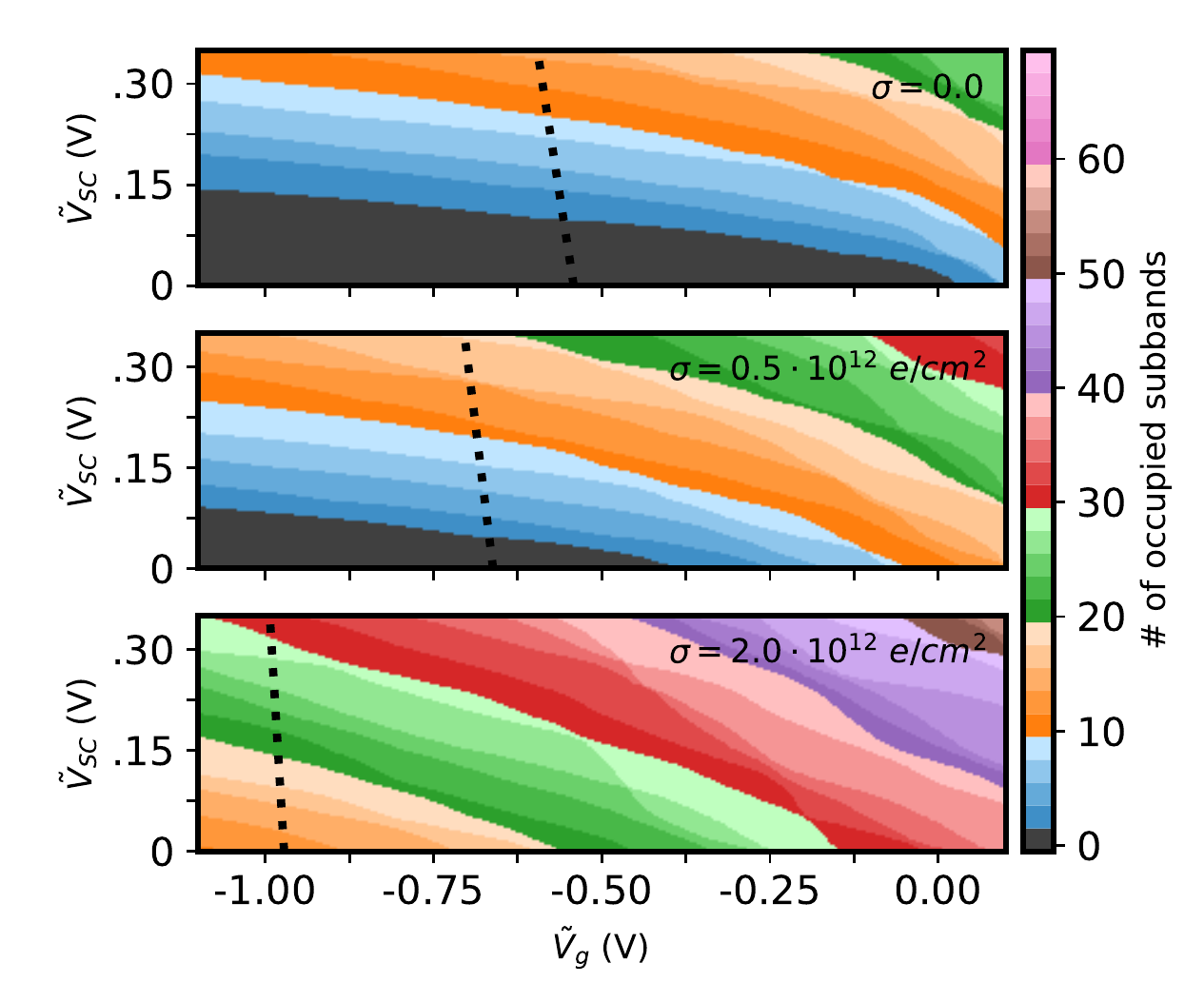}
\end{center}
\vspace{-7mm}
\caption{Subband occupation for a wire of radius $R = 35$ nm as function of the (shifted) gate potential $\widetilde{V}_{g}$ and the (shifted) work function difference $\widetilde{V}_{SC}$. The panels correspond to three different values of surface charge $\sigma$. The negative gate voltage regime  to the left of the dashed lines is characterized by the emergence of holes at the bottom of the nanowire (i.e. near the interface with the dielectric). Note that each spin subband is counted separately (hence, each color/shade corresponds to one pair of (degenerate) spin subbands). }
\label{FIG2}
\vspace{-1mm}
\end{figure}

We note that, for a given surface charge density $\sigma = \rho_{surf} \ell$, the potential $\phi \left(\vec{r}\right)$ depends weakly on the value of $\ell$, as long as $\ell \ll R$. The superconductor is not explicitly included in the Hamiltonian, but it plays a key role in the Poisson problem by setting the Dirichlet boundary condition $\phi \left(\vec{r}\right)=V_{SC}$ at the SM-SC interface, where $V_{SC}$ represents the work function difference between the semiconductor and superconductor \cite{Vuik2016}. The exact value of the work function difference is not known, hence $V_{SC}$ will be treated as a phenomenological model parameter. In this work we consider $V_{SC}> 0$ on the order of $10^2~$mV. Note that a non-zero work function difference results in the bending of the InAs conduction bands near the semiconductor-superconductor interface. We also impose Dirichlet boundary conditions, $\phi \left(\vec{r}\right)=V_{g}$, on the top surface of the back gate. In addition, we impose von Neumann-type boundary conditions on the top, left, and right surfaces of a box of side length $b$ surrounding the wire. We emphasize that the potential within the nanowire is negligibly affected by these boundary conditions -- e.g., the exact value of $b$ or whether we choose von Neumann or Dirichlet boundary conditions -- provided $b \gg R$.  Finally, to reduce the number of parameters in our analysis, we notice that the eigenstates of Eq. (\ref{Ham}) are unaffected by the transformation $\mu \rightarrow  \widetilde{\mu}=0$, $V_{SC} \rightarrow \widetilde{V}_{SC} = V_{SC} + \mu/e$,  $V_{g} \rightarrow \widetilde{V}_{g} = V_{g} + \mu/e$, where $\mu$ is the chemical potential defined with respect to the bottom of the conduction band in a system with $\phi \left(\vec{r}\right)=0$. Our results are given in terms of $\widetilde{V}_{SC}$ and $\widetilde{V}_{g}$ (with $\widetilde{\mu} = 0$); we stress that, in systems with nonzero chemical potential, the values of these parameters are shifted (by $\mu/e$) with respect to the `bare' work function difference and gate potential, respectively. Thus, all one has to do is to redefine $V_{SC}$ and $V_{g}$ by including the chemical potential.

We solve the Schr\"{o}dinger-Poisson equations Eqs. (\ref{Ham}) and (\ref{Pois}) self-consistently using a simple iterative mixing scheme and the finite element package FEniCS \cite{Alnaes2015}. In the calculations {shown in the main text }we use the following parameter values: the effective mass is $m^* = 0.026$ $m_o$ (as we focus on InAs wires; see the appendix for results corresponding to InSb); we consider two different values of the nanowire radius, $R = 35~$nm and $R=50~$nm; the superconductor and dielectric thicknesses are $t = 10~$nm and $d = 10~$nm, respectively; the dielectric constants of the wire, dielectric, and surrounding air are taken to be $\epsilon_{wire} = 15.15 \epsilon_o$, $\epsilon_{diel} = 24 \epsilon_o$, and $\epsilon_{air} = \epsilon_o$, where  $\epsilon_o$ is the permittivity of free space. 

\section{Results}

\begin{figure}[t]
\begin{center}
\includegraphics[width=0.48\textwidth]{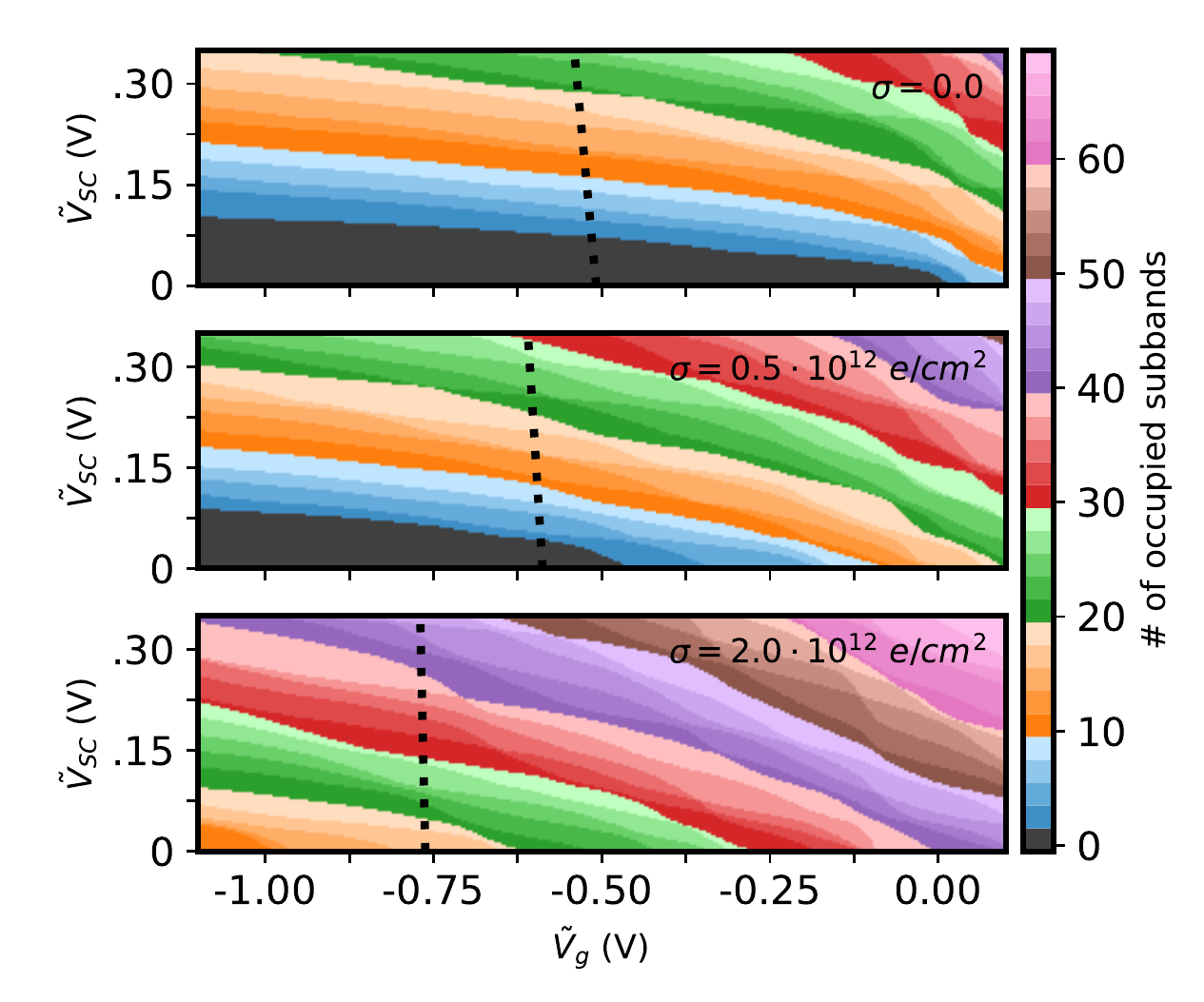}
\end{center}
\vspace{-7mm}
\caption{Same as Fig. \ref{FIG2} for a wire of radius $R=50~$nm. Note that the number of occupied subbands increases with the radius of the wire.}
\label{FIG3}
\vspace{-1mm}
\end{figure}

We first investigate the dependence of the  subband occupation on the (shifted) gate and superconductor potentials for a wire of radius $R = 35~$nm.
The results are shown in Fig. \ref{FIG2}. The three panels correspond to three different values for the surface charge density $\sigma = \rho_{surf} \ell$. Note that spin subbands are counted separately, so each change of color/shade represents the occupation of an additional pair of (degenerate) spin subbands. We focus on negative gate voltages since we are interested in limiting the number of occupied subbands. Note that, in addition to increasing the number of occupied subbands, attractive (i.e. positive) gate potentials will move the conduction electrons away from the SM-SC interface and, consequently, suppress the superconducting proximity effect.

The subband occupation in the absence of surface charge ($\sigma =0$) is shown in Fig. \ref{FIG2}(a). As expected, there are no conduction electrons in the wire for $\widetilde{V}_{SC} = 0$ and $\widetilde{V}_{g}\leq 0$, as the chemical potential is below the bottom of the conduction band.  Increasing 
$\widetilde{V}_{SC}$ (for a fixed value of the gate potential) bends the conduction band near the SM-SC interface resulting in an increasing number of 
occupied subbands (up to about 26 for $\widetilde{V}_{g}=0$ and $\widetilde{V}_{SC} =325~$mV).  Of course, in the actual system 
$\widetilde{V}_{SC}$ is fixed and the available tunning parameter is the gate potential $\widetilde{V}_{g}$. An important feature revealed by our results is the possibility of tunning  across several subband pairs using moderate gate voltages. 
Note that this property is independent of the work function difference, provided $\widetilde{V}_{SC}$ is large enough so that the system is not nearly depleted at  $\widetilde{V}_{g} = 0$ (i.e. $\widetilde{V}_{SC}>100~$mV). The possibility of observing these subband crossings experimentally will be discussed below. 

\begin{figure}[t]
\begin{center}
\includegraphics[width=0.48\textwidth]{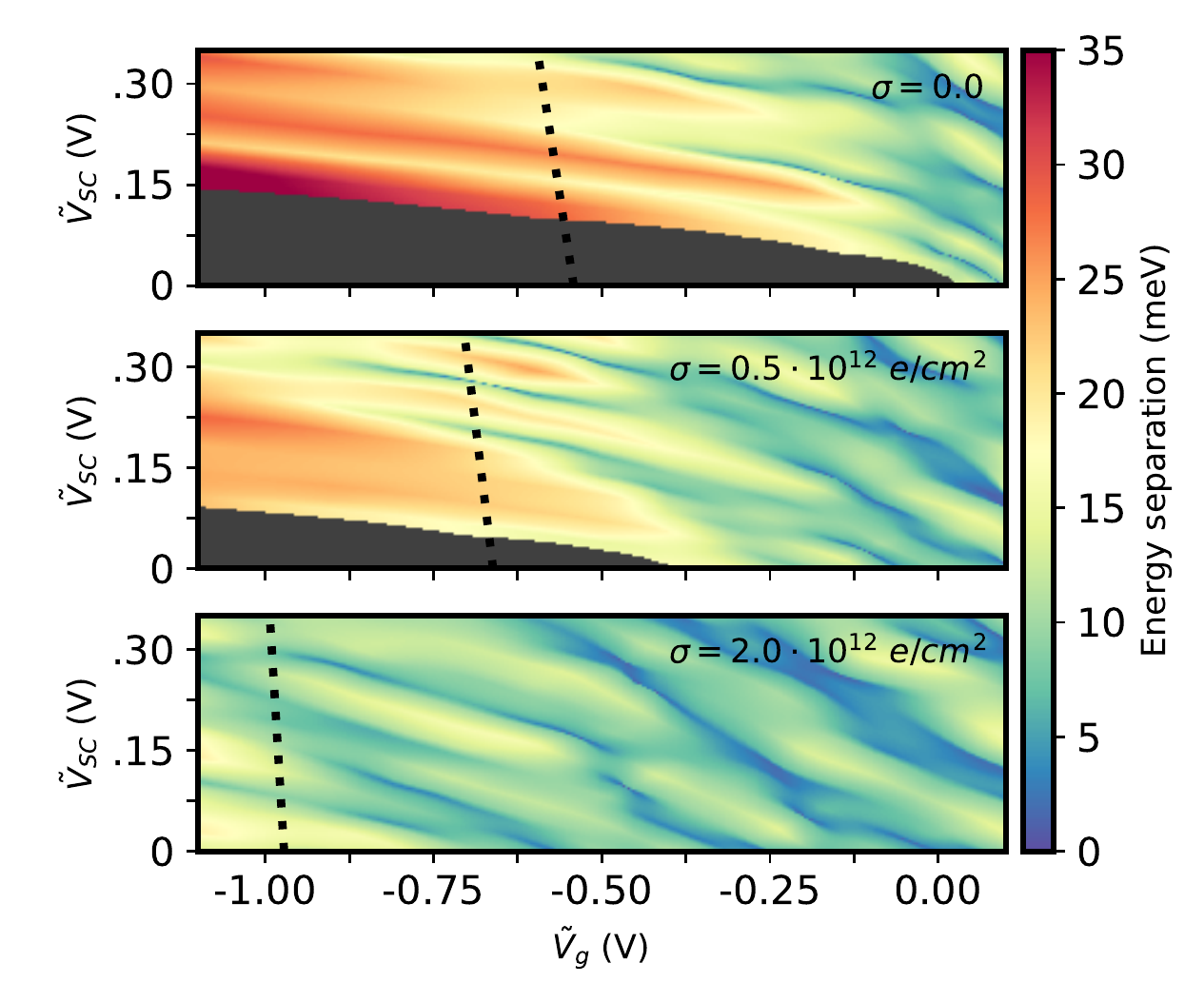}
\end{center}
\vspace{-7mm}
\caption{Average subband separation near the Fermi level, as defined by Eq. (\ref{delE}), for a wire of radius $R=35$ nm as function of  gate voltage, $\widetilde{V}_{g}$, and SM-SC work function difference, $\widetilde{V}_{SC}$. The panels correspond to three different values of surface charge, $\sigma$, while the dashed lines indicate the emergence of holes.}
\label{FIG4}
\vspace{-1mm}
\end{figure}

\begin{figure}[t]
\begin{center}
\includegraphics[width=0.48\textwidth]{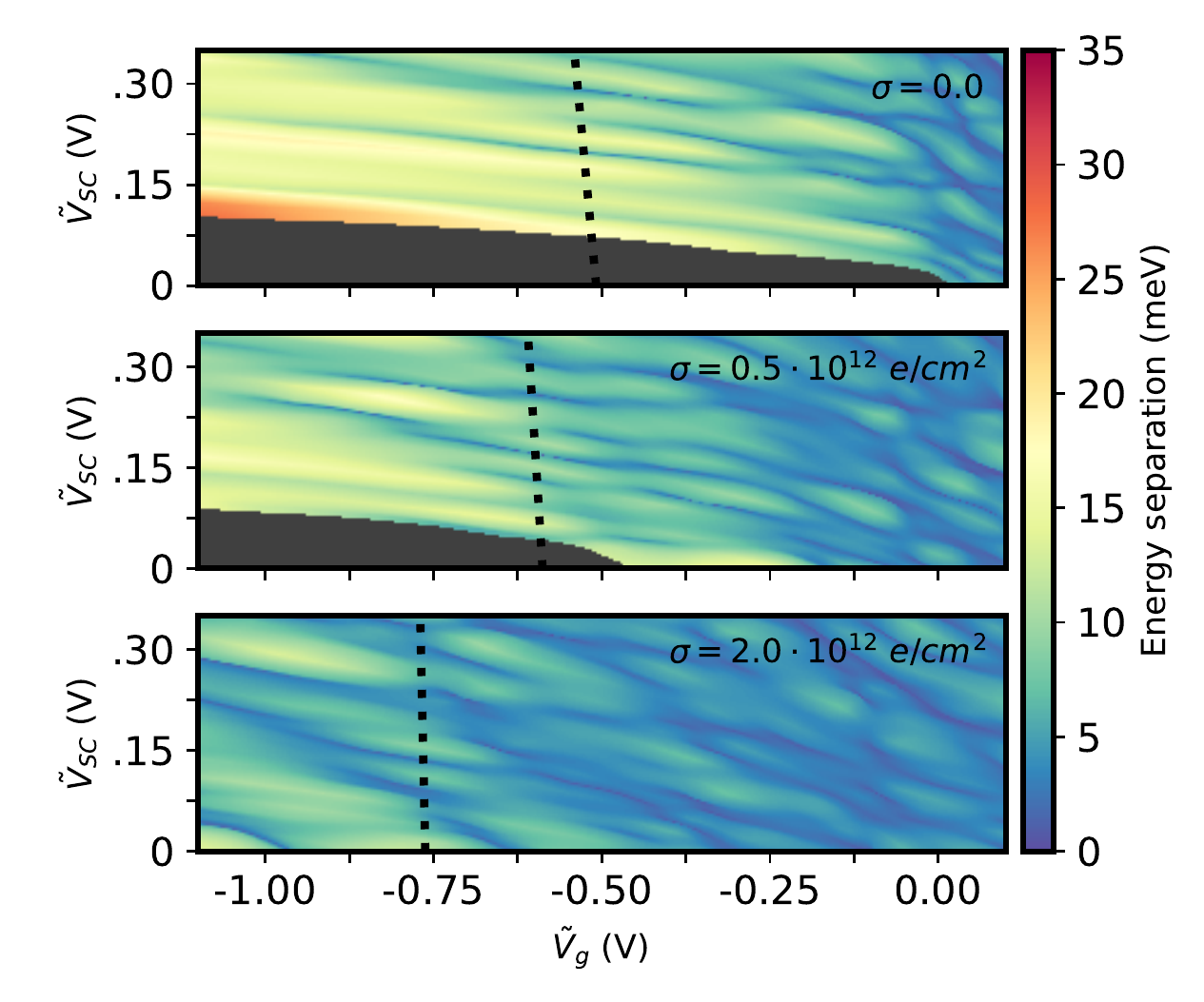}
\end{center}
\vspace{-7mm}
\caption{Same as Fig. \ref{FIG4} for a wire of radius $R=50~$nm. Note that higher subband occupation (see Figs. \ref{FIG2} and \ref{FIG3}) is consistently associated with smaller inter-band separation.}
\label{FIG5}
\vspace{-1mm}
\end{figure}

A crucial question is whether single-subband or few-subbands occupancy can be obtained by applying a negative-enough gate potential. The relevance of this question becomes clear in the context of realizing topological superconductivity and Majorana zero modes (MZMs) using hybrid SM-SC structures, as high occupancy is not only detrimental to the stability of MZMs (as a result of being associated with smaller topological gaps), but also generates ubiquitous trivial low-energy states in the presence of disorder or system inhomogeneities \cite{Bagrets2012,Liu2012,DeGottardi2013,Rainis2013,Adagideli2014,Pikulin2012,Woods2019b,Chen2019}. To answer this question, we first notice that, in principle, the conduction bands will eventually become completely depleted for a large enough value of $|\widetilde{V}_{g}|$. However, it is possible that the valence bands are also partially depleted (i.e. holes emerge) before reaching this regime. The emergent holes will be localized near the interface with the dielectric (i.e. as close as possible to the back gate) and, consequently, will practically experience no proximity  from the parent superconductor. As a result, a SM-SC hybrid system in this regime will consist of a quasi-1D topological superconductor (made of proximitized conduction electrons) and a parallel `normal-metal' channel (made of holes).  Of course, the hole channel will, in general, completely destroy the topological properties of the MZMs supported by the parallel 1D superconductor; hence, its emergence has to be avoided. To account for the possible emergence of holes, we estimate the corresponding  $\widetilde{V}_{g}$ values using a Luttinger model for the InAs light and heavy hole bands, considering them as decoupled from the conduction electrons. The gap between the conduction and valence band edges is $E_{gap} =  0.418$ eV \cite{Luttinger1956,Winkler2003} (see Appendix (\ref{Lut}) for more details). The dashed black lines in Figs. \ref{FIG2}-\ref{FIG5} indicate the critical (shifted) gate voltage associated with the emergence of  holes. Note that the electrostatic screening from the holes (which is significant, particularly considering their higher effective mass) has not been explicitly taken into account in our calculations. This implies that the actual lever arm associated with $\widetilde{V}_{g}$ is reduced within the hole-regime left of the dashed lines. Practically, from the perspective of Majorana physics, the dashed lines mark the limit of negative (shifted) potentials consistent with the realization of topologically-protected MZMs. 
For example, Fig. \ref{FIG2}(a) shows that holes begin to emerge near $\widetilde{V}_{g} \approx -550~$mV. This limits drastically the window of $\widetilde{V}_{SC}$ values associated with the few-subband regime, $\widetilde{V}_{SC} < 200~$mV. This window is even smaller in panel (b) and absent in panel (c).    

The impact of surface charge (which is responsible for band-bending near the surface of the SM wire) on the subband occupation is illustrated in Fig. \ref{FIG2}(b) and (c). The chosen surface charge values are consistent with experimental estimates for InAs \cite{Olsson1996} and have been used in previous theoretical studies of  Majorana devices \cite{Winkler2019,Escribano2019}. As expected, the main effect of the (positive) surface charge is to increase the overall number of occupied subbands. The effect  can be clearly  seen in panels (b) and (c), as the subband occupations are shifted down and to the left relative to panel (a). Note that, despite the fact that the gate potentials for which holes begin to populate the wire become more negative as $\sigma$ increases, the window of $\widetilde{V}_{SC}$ values associated with the few-subband regime shrinks and eventually vanishes, as illustrated in panel (c). For example, the single-subband regime can be attained in Fig. \ref{FIG2}(b) only for $\widetilde{V}_{SC} \lesssim 75~$mV, while in Fig. \ref{FIG2}(c) the fewest number of occupied subbands is ten (corresponding to $\widetilde{V}_{SC}=0$). 

In addition to  $\widetilde{V}_{SC}$  and  $\widetilde{V}_{g}$, the subband occupation has a significant dependence on the radius of the wire. To illustrate this feature, in  Fig. \ref{FIG3} we shows the subband occupation for the same  $\widetilde{V}_{SC}-\widetilde{V}_{g}$ parameter ranges as Fig. \ref{FIG2}, but for a thicker wire with $R = 50~$nm. On the one hand, wires with larger diameters may be less prone to disorder, as compared to thinner wires, but, on the other hand, the reduced confinement implies a smaller inter-subband energy spacing. This leads to a higher subband occupancy for any given set of potential parameters, as manifest from the  comparison of  Figs. \ref{FIG2} and \ref{FIG3}. 
 Note that the $R = 35~$nm and $R=50~$nm wires generate qualitatively similar trends, with the thicker wire being associated with higher occupancies. It is still possible to tune the system through several subbands before the hole states become populated. However, the gate voltages associated with the emergence of holes  shift with increasing $R$ toward less negative values, due to the combined effects of reduced confinement, reduced electrostatic screening, and larger separation between the back gate and the superconductor. This implies that, if attaining the few-subband regime (ideal for realizing MZMs) is not possible by gating (due to the emergence of holes), reducing the diameter of the wire is a possible solution. 

The fundamental reason why high subband occupancy is detrimental to realizing  Majorana bound states is that the corresponding (typical) value of the inter-subband energy spacing for subbands near the Fermi energy is significantly reduced as compared with the few-subband regime. On the one hand, this is due to a weaker transverse confinement of the high-energy subbands (as compared to the low-energy subbands) by the effective potential $\phi \left(\vec{r}\right)$ and, on the other hand, to a larger effective mass associated with  high-energy subbands (which is an effect of the non-parabolicity of the conduction band). Note that our effective mass approximation does not capture the second effect, which requires a multi-orbital modeling of the semiconductor nanowire \cite{Woods2019a,Winkler2003}. The reduced inter-subband energy separation associated with high subband occupancy makes the system susceptible to perturbations such as  disorder and inhomogeneous effective potentials due to the presence of multiple gates, strain, etc. If the subbands are not well separated, these perturbations  lead to inter-subband mixing \cite{Woods2019b, Moor2018}, which, in turn, induces (topologically-trivial) low-energy Andreev bound states (ABSs) \cite{Chen2019}. Note that this type of near-zero energy states are generic in class D systems \cite{Pan2019,Beenakker2015} and were explicitly shown to emerge in proximitized nanowires (in the presence of spin-orbit coupling and nonzero Zeeman field) within a large window of local gate potentials \cite{Woods2019b,Chen2019}.  We emphasize that the emergence of this type of low-energy states does not require the explicit presence of disorder, as multiple subband occupancy leads to an effective random matrix Hamiltonian in the presence of any perturbation that couples the subbands (e.g., applied gate potentials, strain, etc.). Finally, note that certain types of trivial low-energy states -- the so-called quasi-Majorana \cite{Vuik2019}, or partially-separated ABS states \cite{Moore2018} -- mimic very faithfully the local phenomenology of Majorana zero modes \cite{Moore2018a,Stanescu2018b}. 

To evaluate the dependence of the inter-subband energy separation on the relevant system parameters, we calculate the average subband energy separation near the Fermi level as functions of  the (shifted) gate potential $\widetilde{V}_{g}$ and the (shifted) work function difference $\widetilde{V}_{SC}$ for two wires with radii $R = 35~$nm and $R = 50~$nm. The results are shown in  Figs. \ref{FIG4} and \ref{FIG5}, respectively. 
The average subband separation near the Fermi level is defined as follows. Let  the $n^{th}$ subband be the `Fermi subband', i.e. the subband having its bottom closest to the Fermi level. The energy at the bottom of the Fermi subband is $E_n$. We define the average subband separation as the weighted average of the gaps between the Fermi subband and the subbands immetiadely below and above it. Explicitly, we have
\begin{equation}
\left(\Delta E\right)_{ave} = \lambda \left(E_n - E_{n-1}\right) + (1 - \lambda) \left(E_{n+1} - E_{n}\right),  \label{delE}
\end{equation}
where $\lambda = \left(E_{n+1} + E_n\right)/\left(E_{n+1}-E_{n-1}\right)$. We first point out that the average energy separation tends to increase as $\widetilde{V}_{g}$ becomes more negative. This effect is due to the increased confinement generated by the negative gate potential, which pushes the electron wave function towards the semiconductor-superconductor interface, combined with the reduction of subband occupation. By contrast,  increasing $\widetilde{V}_{SC}$, while also enhancing the effective confining potential, increases the subband occupation as well, the net effect on the subband energy separation being minimal. 

The results shown in Figs. \ref{FIG4} and \ref{FIG5} suggest that the optimal system for realizing robust MZMs would be a thin wire with no surface charge (or a small $\sigma$), as the corresponding spectrum is characterized by a large inter-subband energy separation for moderate negative gate potentials, before the emergence of holes in the system. For example, the zero surface charge results in Fig. \ref{FIG4}(a) present an excellent outlook for the realization of MZMs, as the Fermi level subband separation (right before the onset of hole states) is $\left(\Delta E\right)_{ave} \approx 15 - 30~$meV, depending upon the exact value of $\widetilde{V}_{SC}$. The situation becomes progressively less promising with the addition of surface charge, as can be seen in Figs \ref{FIG4} (b) and (c).  This is due to the reduced confinement caused by the positive surface charge pulling the wave functions away from the semiconductor-superconductor interface.  Increasing the diameter of the wire further reduces the confinement and, consequently, the subband energy separation, as manifest when comparing the results in Fig. \ref{FIG4} and Fig. \ref{FIG5}. 
In addition, in thicker wires the hole regime sets in at less negative gate potentials, making the small inter-subband energy separation  problem even worse. In Fig. \ref{FIG5}(c), for example, the subband energy separation is $\left(\Delta E\right)_{ave} \approx 2 - 7$ meV for all acceptable values of the gate voltage (i.e. outside the hole regime) due to the combined effects of large surface charge and large wire diameter.  As an estimate of the relative impact of these factors, we note that the average subband energy separations in Figs. \ref{FIG4}(c) (large $\sigma$, small $R$) and \ref{FIG5}(a) ($\sigma=0$, large $R$)  are comparable. This points to the importance of using wires with the smallest possible diameter in order to keep the subband occupancy a minimum. Finally, we note that the effective mass model used here predicts larger subband separation energies than more detailed (multi-orbital) models, which take into account the nonparabolicity of the conduction band \cite{Winkler2003,Stanescu2013}. Hence, the actual subband separation is expected to  be smaller than the predictions based on the results shown in Figs. \ref{FIG4} and \ref{FIG5}. These nonparabolic effects are larger for high-energy subbands and will generate an additional reduction of the inter-band separation in the high-occupation regime. Thus, our calculations presented here provide the most optimistic estimates for inter-subband energy separations with the corresponding realistic level separations in experimental nanowires likely being even smaller.

Our results show clearly that i) regimes characterized by many occupied subbands are possible and even likely within realistic windows of system parameters, ii) gating the system to reach the few-subbands regime is usually not possible (except narrow parameter windows) due to the onset of holes, and iii) the many-subbands regime is typically associated with small values of the inter-subband energy spacing near the Fermi level, which has highly detrimental consequences for the realization and observation of robust Majorana zero modes. The natural question is how to measure experimentally the subband occupation of a proximitized wire? This would facilitate the optimization of the growth and fabrication procedures in order to reach the `ideal' few-subbands regime. Our key observation is that the most striking feature associated with a quasi-1D subband becoming occupied is a van Hove singularity in the density of states (DOS), or the local DOS (LDOS), at the Fermi energy. This is the well-known square-root van Hove singularities associated with 1D subbands in semiconductor quatum wires pointed out more than thirty years ago \cite{Sarma1987}. Probes that are highly sensitive to this van Hove singularity would be good candidates for experimentally estimating the subband occupation.

A straightforward and powerful tool for measuring the local density of states is scanning tunneling spectroscopy \cite{Binnig1987}. To anticipate the capabilities and limitations of such a tool, we calculate the LDOS on different facets of a proximitized nanowire with a transverse profile as shown in Fig. \ref{FIG1}. 
More specifically, we assume that the system is in the normal state (i.e. above the superconducting critical temperature) and that the SM-SC coupling is not too high (i.e. it is not much larger than the gap of the parent superconductor) and  focus on the LDOS at zero energy as a function the applied gate potential $\widetilde{V}_{g}$. As the applied potential becomes more negative, the occupied SM conduction subbands are depleted, the presence of a subband bottom near the Fermi energy being signaled by a peak in the LDOS. The results for a wire of radius  $R=35~$nm and different representative combinations of $\widetilde{V}_{SC}$ and $\sigma$ values are shown in Fig. \ref{FIG6} (see Appendix \ref{LDOS} for technical details). The blue, red, and green lines correspond to the LDOS on the three exposed facets of the wire (which are potentially accessible to an STM measurement), as indicated by the corresponding colored dots in the inset of Fig \ref{FIG6}(a). The dashed lines indicate the $\widetilde{V}_{g}$ values corresponding to the bottom of a pair of (degenerate) spin subbands crossing the Fermi level, while the numbers indicate the index of the corresponding pair.

The first significant feature is the presence of well-defined peaks in the LDOS (due to van Hove singularities) that coincide with subband bottoms crossing the Fermi energy (dashed lines). Note that the peaks measured on different facets that are associated with the same subband pair occur at exactly the same value of $\widetilde{V}_{g}$, but may have significantly different amplitudes. These amplitudes depend on the transverse profile of the subband, which, in general, are different on different facets of the wire. Moreover, the lowest energy subbands are 'pushed away' by the negative gate potential and strongly confined near the SM-SC interface. Consequently, they have negligible amplitudes on the exposed facets of the wire and are practically 'invisible' in a LDOS measurement.   For example, the first subband pair in Fig. \ref{FIG6}(a) has no visible peak in the LDOS, while the second pair generates a significant signal only on the "green" facet (i.e. close to the SM-SC interface and as far as possible from the potential gate).  The existence of 'invisible' subbands prohibits a simple counting of the subband occupation number,  as the number of observed LDOS peaks is, in general, lower. In other words, the number of observed LDOS peaks as function of the applied gate potential (obtained, for example, in an STM experiment) is typically lower than the number of occupied subbands.

\begin{figure}[t]
\begin{center}
\includegraphics[width=0.48\textwidth]{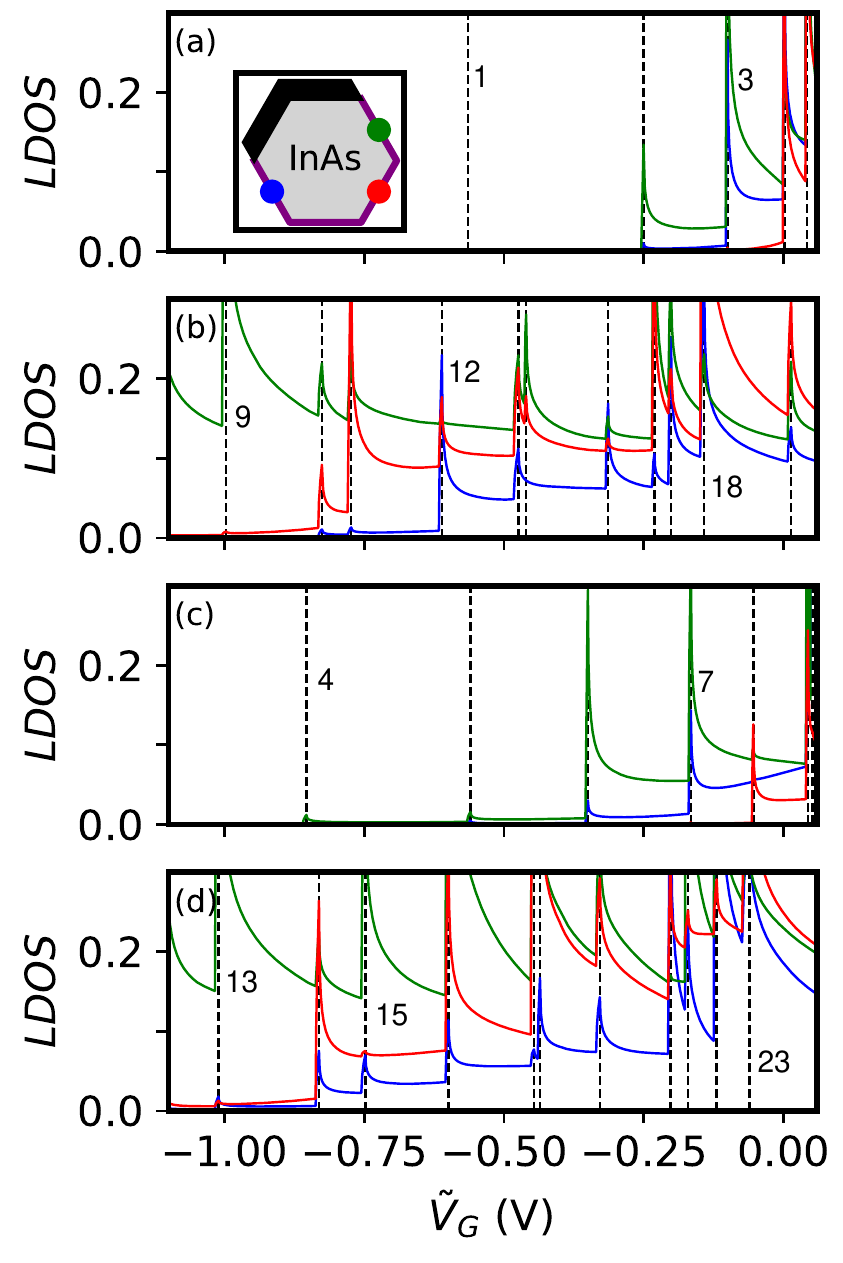}
\end{center}
\vspace{-7mm}
\caption{Local density of states at zero energy versus (shifted) gate voltage $\widetilde{V}_{g}$ for a wire of radius $R=35$ nm. The LDOS is normalized with respect to the maximum in each panel. The three sets of lines correspond to the local density of states in the regions marked by the corresponding colored dots in the insert of panel (a). The dashed lines indicate the  $\widetilde{V}_{g}$ values where the bottom of a pair of spin subbands cross the Fermi level. Numbers next to dashed lines indicate subband pair index. The model parameters corresponding to different panels are: (a) $\widetilde{V}_{SC} = 0.1 V$, $\sigma = 0$, (b) $\widetilde{V}_{SC} = 0.1 V$, $\sigma = 2\cdot 10^{12}$ e$\cdot$ cm\textsuperscript{2}, (c) $\widetilde{V}_{SC} = 0.25 V$, $\sigma = 0$, (d) $\widetilde{V}_{SC} = 0.25 V$, $\sigma = 2\cdot 10^{12}$ e$\cdot$ cm\textsuperscript{2}. }
\label{FIG6}
\vspace{-1mm}
\end{figure}

However, the very observation of a peaked LDOS structure would have enormous implications and would contain significant information regarding the hybrid system and its suitability for realizing Majorana physics. First, the presence of sharp LDOS peaks signals the existence of well defined subbands within the nanowire. Disorder will generically broaden these peaks, as states become localized within different regions of the wire \cite{Sarma1987}. In the presence of strong disorder, the peaked structure is washed out, the notion of subband becomes meaningless, and the possibility of realizing Majorana zero modes is practically lost. It is therefore essential that the nanowire mobility in the SM-SC hybrid is large so that any disorder-induced collisional broadening is smaller than the inter-subband energy separation for the hybrid system to be able to host MZMs \cite{Sarma1987}. Moreover, even if a peaked structure can be observed, one has to check its independence on the position along the wire. Basically, a structure dependent of the position along the wire is a clear indication of long-range effective potential variations, a scenario likely to lead to the emergence of partially-separated ABSs (i.e. quasi Majoranas), rather than well-separated MZMs. Finally, each identification of a LDOS peak (corresponding to the bottom of a subband being at the Fermi level) provides a value of the gate potential at which Majorana physics should be explored. In other words, if the system can support MZMs (i.e. if it is uniform and clean enough), the optimal gate voltage regimes for the realization of these MZMs will necessarily correspond to the chemical potential being close to the bottom of a confinement-induced subband, which is indicated by the LDOS peak. 

\begin{figure}[t]
\begin{center}
\includegraphics[width=0.48\textwidth]{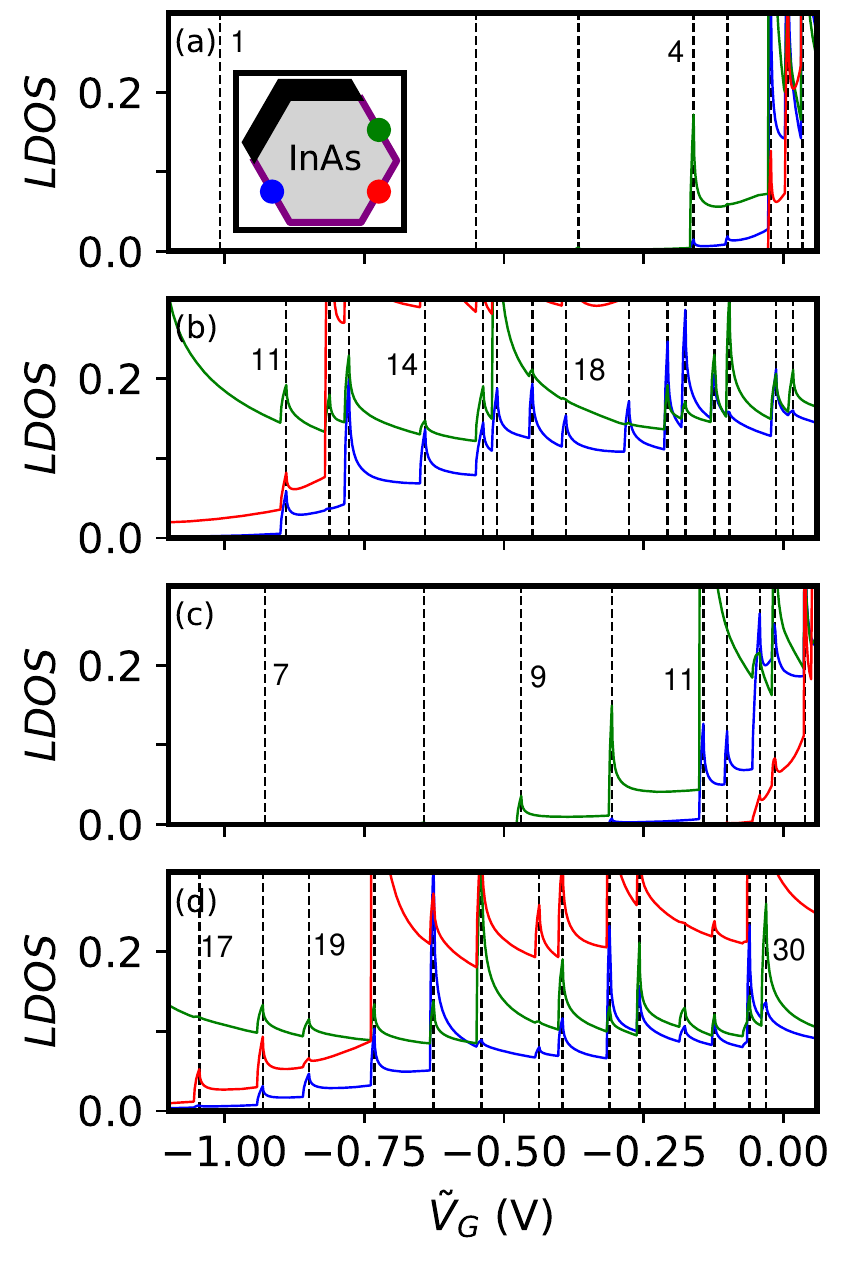}
\end{center}
\vspace{-7mm}
\caption{Same as Fig. \ref{FIG6} for a wire of radius $R=50~$nm. Note that the ranges of gate potentials over which LDOS peaks are visible are similar to those in Fig. \ref{FIG6}, but there are more 'invisible' subbands.}
\label{FIG7}
\vspace{-1mm}
\end{figure}

The details of the LDOS structures shown in Figs. \ref{FIG6} (wire of radius $R=35~$nm) and \ref{FIG7} (wire of radius $R=50~$nm) reveal additional information about the system. For example, let us compare a wire with no surface charge, $\sigma = 0$, panels (a) and (c), and a wire with large surface charge density, $\sigma = 2 \cdot 10^{12} \> e/{\rm cm}^2$, panels (b) and (d). In the absence of surface charge the LDOS for all three facets shows a peak structure that diminishes rapidly as $\widetilde{V}_{g}$ become more negative and the wave functions become strongly confined near the SM-SC interface. 
By contrast, the LDOS peaks for the wire with large surface charge remain moderately large across the whole scan of $\widetilde{V}_{g}$. In particular, the LDOS peaks in the upper right facet region (green curve) is quite large. The qualitative differences between the two cases are determined by the attractive potential associated with the positive surface charge, which keeps a significant fraction of the wave function weight near the exposed edges making them 'visible' in the LDOS, while the negative gate potential tends to push the states towards the semiconductor-superconductor interface. The upper right facet maintains especially prominent peaks because of its larger spatial separation from the repulsive back-gate. Note that the observation of multiple LDOS peaks -- as, e.g., in panels (b) and (d) -- is a clear indication of high subband occupancy, a regime that is not favorable to the realization of stable MZMs, as argued above. Also, the observation of only a few peaks over a moderate range of gate potentials is not necessarily an indication that the system is in the few-subband regime. However, combined with a reduction of the wire diameter, such an observation indicates a clear path toward the few-subbands regime. Note that the qualitative (and even some of the semi-quantitative) features are very similar in the corresponding panels of Figs. \ref{FIG6} and  \ref{FIG7}. Assume, for example, that the system is in a regime corresponding to panels (c), i.e., $\widetilde{V}_{SC} = 0.25 V$, $\sigma = 0$, and that the lowest observed peaks occur around  $\widetilde{V}_{g} \approx -0.3 V$. In the thinner wire [Fig. \ref{FIG6}(c)] the peak corresponds to the $n=6$ subband pair (i.e. 12 occupied spin sub-bads), while in the thicker wire [Fig. \ref{FIG7}(c)] the peak corresponds to the $n=10$ subband pair (i.e. 20 occupied spin sub-bads). All other conditions being equal, the thinner wire is clearly more suitable for realizing stable MZMs. Finally, we emphasize that i) the LDOS measured in the normal states (i.e. above the critical superconducting temperature) already contains a wealth of information about the feasibility of Majorana physics in a SM-SC hybrid system and, should experimental LDOS data become available (e.g., from future STS measurements on SM-SC structures), ii) it is essential to perform detailed numerical studies (using realistic models) of the dependence of the LDOS features on key factors, such as disorder and long range potential inhomogeneities. We emphasize that experimental studies of nanowire LDOS manifesting clear 1D van Hove singularities at the subband bottoms (which necessitate that the inter-subband energy separations are larger than disorder broadening) along with the occupancy of only a few (ideally just one) orbital subband(s) are essential steps in advancing the search for MZMs in SM-SC hybrid systems.

\section{Summary}

We have studied the subband occupation of proximitized InAs  nanowires as a function of the applied gate potential and the semiconductor-superconductor work function difference in the presence of surface charges using an effective mass model and  solving the corresponding Schr\"{o}dinger-Poisson equations self-consistently.  We find that for realistic values of the surface charge density the system is characterized by many occupied subbands, while reducing the occupation and driving the system into a few subband regime using a (negative) gate potential is precluded by the emergence of holes in the semiconductor valence band, {unless the system is characterized by a small work function difference $\widetilde{V}_{SC}$}.
The many subband regime is characterized by small inter-subband energy separation near the Fermi level. The small subband energy separation, in combination with the sheer number of occupied subbands, makes the system susceptible to  subband mixing in the presence of disorder, potential inhomogeneities, strain, and other perturbations, which generically produces low-energy Andreev bound states. 
We emphasize that, even in the absence of disorder, a many subband system with nonzero Zeeman field, spin-orbit coupling, and induced superconductivity effectively behaves as a class D system due to the practically uncontrollable inter-subband coupling that occurs once translational symmetry is broken. Consequently, topologically-trivial low-energy states will emerge generically in such a system as the control parameters (e.g., gate potentials and Zeeman fields) are tuned. Therefore, the many subband regime should be systematically avoided to ensure the realization of stable topological superconductivity and Majorana zero modes. In other words, in addition to the commonly recognized conditions associated with the realization of topologically-protected MZMs -- such as having long wires and the  absence of disorder -- having low-occupancy, i.e., chemical potential values consistent with one or a few occupied subbands, represents an essential requirement. We note that using a semiconductor with a lower effective mass, e.g., InSb, does not automatically solve the problem, as illustrated by the results shown in Appendix \ref{InSb}. A smaller gap leads, on the one hand, to the emergence of holes at less negative values of the gate potential, hindering the depletion of the wire. On the other hand, a smaller gap enhances the non-parabolicity effects (not included in our effective mass model), which results in a further reduction of the inter-subband separation at high occupancy. 

All these considerations makes it clear that measuring the subband occupancy and, ultimately, being able to reduce it to the few-subbands regime are critical tasks for building Majorana devices.  To identify experimentally the systems which may be in the many subband regime and to be able to optimize the growth and fabrication procedures, we propose a measurement of the LDOS on the exposed facets of the nanowire, e.g., using an STM. We note that such a measurement is not capable, in general, to give the exact number of occupied subbands, as states strongly localized near the SM-SC interface may remain 'invisible'. On the other hand, such a measurement can provide critical information regarding the suitability of a SM-SC hybrid structure to host Majorana physics, even when performed in the normal state (i.e. above the critical superconducting temperature). In particular, the observation of peaks in the LDOS as function of the applied gate voltage indicates that the system is clean-enough to have some well-defined subbands. Furthermore, the positions of the peaks provide the values of the gate voltage where Majorana physics should most likely emerge. The presence of disorder is expected to destroy the LDOS peaks, while a LDOS peak structure dependent on the spatial position along the wire may be an indication of long-range effective potential variations, which may generate quasi-Majorana modes. If LDOS peaks are completely absent in the nanowires, the system is simply unsuitable for studying Majorana physics because disorder would then completely suppress the emergent Majorana modes.  Even if the LDOS peaks are seen, one must worry about the subband occupancy question, as discussed in depth in our work.

This work is supported by the Laboratory for Physical Sciences and Microsoft.

\appendix

\section {Luttinger model for hole states} \label {Lut}
The valence band of the InAs nanowire is modeled with a four band Luttinger model, that incorporates the degenerate light and heavy hole bands characteristic of the valence band edge of III-V semiconductors \cite{Luttinger1956,Winkler2003}. We take the wire to be grown along the $\left[111\right]$ direction. The Hamiltonian is given by 
\begin{equation}
H_{Lutt} = 
\begin{bmatrix}
-P - Q & S & R & 0 \\
S^\dagger & -P + Q & 0 & R \\
R^\dagger & 0 & -P + Q & -S \\
0 & R^\dagger & -S^\dagger & -P-Q
\end{bmatrix},
\end{equation}
where 
\begin{align}
P &= \gamma_1\frac{\hbar^2 k^2}{2m} + E_{gap} + \mu + e\phi(\vec{r}), \\ \label {A2}
Q &= \gamma_3 \frac{\hbar^2}{2m}\left(k_x^2 + k_y^2 - 2k_z^2\right), \\
S &= \frac{2}{\sqrt{3}}(2\gamma_2 + \gamma_3) \frac{\hbar^2}{2m} \left(k_x - i k_y\right)k_z, \\
R &= \frac{1}{\sqrt{3}}(\gamma_2 + 2\gamma_3) \frac{\hbar^2}{2m} \left(k_x - i k_y\right)^2, \label {A5}
\end{align}
and we have employed the axial approximation for simplicity \cite{Winkler2003}.
The $\gamma_i$ coefficients are known as the Luttinger parameters and take the values $\gamma_1 = 20.4$, $\gamma_2 = 8.3$, and $\gamma_3 = 9.1$ for InAs. The valence band is separated from the conduction band by a bandgap, given by $E_{gap} = 0.418$ eV for InAs, and we have incorporated the chemical potential $\mu$ and electrostatic potential $\phi$ into the diagonal term $P$. When the model is implemented on the nanowire, we take $k_x \rightarrow - i \partial_x$,  $k_y \rightarrow - i \partial_y$, and use finite element method \cite{Alnaes2015} to solve for the eigenstates near the valence band edge. As mentioned in the main text, the charge density from populated hole states is not included in the electrostatics calculations. Rather, we are interested in the $\widetilde{V}_g$ values for which the hole states begin to populate.

\begin{figure}[t]
\begin{center}
\includegraphics[width=0.48\textwidth]{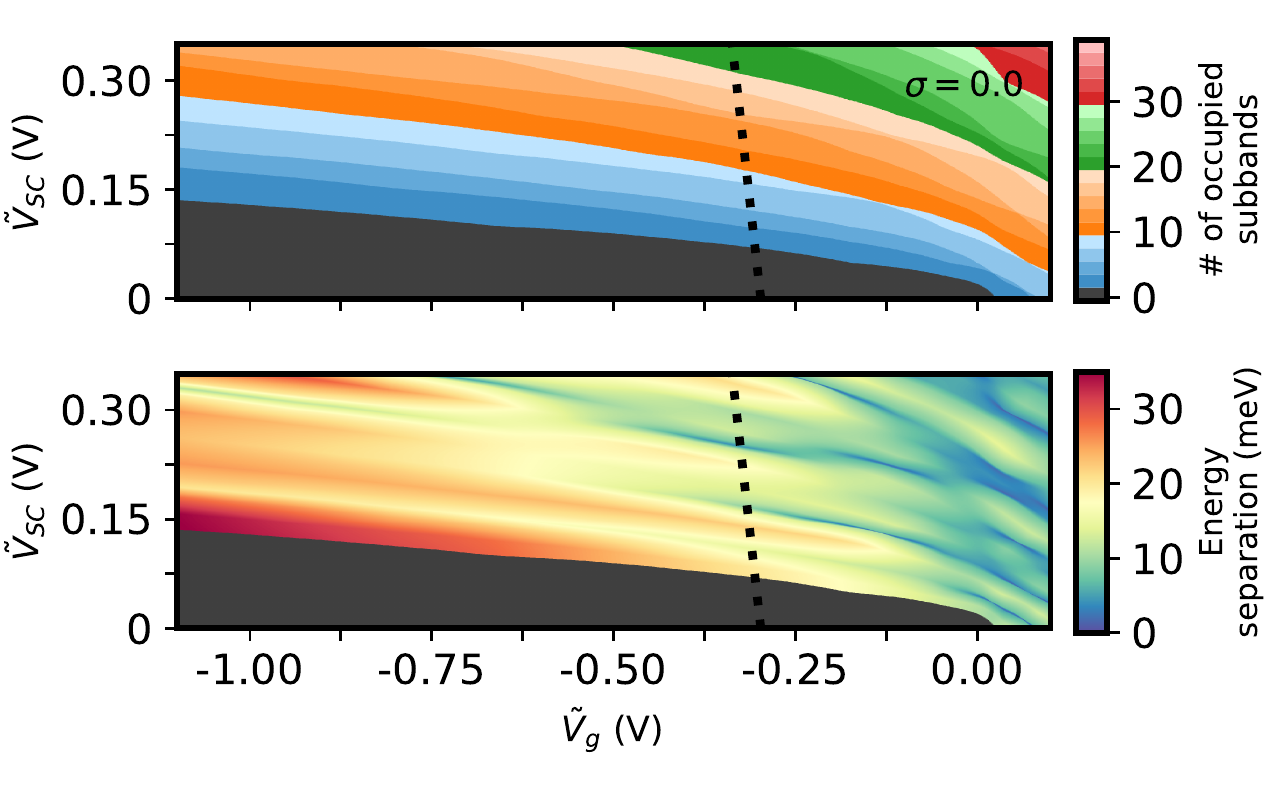}
\end{center}
\vspace{-7mm}
\caption{(a) Subband occupation for an InSb wire of radius $R = 50$ nm as function of the (shifted) gate potential $\widetilde{V}_{g}$ and the (shifted) work function difference $\widetilde{V}_{SC}$. The negative gate voltage regime  to the left of the dashed lines is characterized by the emergence of holes at the bottom of the nanowire (i.e. near the interface with the dielectric). (b) Average subband separation near the Fermi level, as defined by Eq. (\ref{delE}).}
\label{FIG8}
\vspace{-1mm}
\end{figure}

\section{Subband occupation and energy separation in InSb nanowires} \label{InSb}
InSb, another III-V semiconductor often used in the fabrication of Majorana devices, is characterized by an effective mass significantly lower than InAs. To asses the consequences of having a lower effective mass,  we provide a calculation of the subband occupation and energy separation for InSb similar to the calculation done for InAs in the main text. We use the same device setup and values of the material parameters corresponding to InSb: $m^* = 0.015$ $m_o$ and $\epsilon_{wire} = 16.8\epsilon_o$. The materials parameters for the corresponding Luttinger model are $\gamma_1 = 34.8$, $\gamma_2 = 15.5$, $\gamma_3 = 16.5$, and $E_{gap} = 0.235$ eV \cite{Winkler2003}. InSb does not have a large surface charge density, like InAs, so we set $\sigma = 0$. The subband occupation and average separation near the Fermi level  for an InSb nanowire of radius $R = 50$ nm
are shown in Fig. \ref{FIG8} (a) and (b), respectively. Comparing the InSb subband occupation to that of the InAs wire in Fig. \ref{FIG3}(a), we notice that the InSb wire has a slightly fewer number of occupied subbands for a given set of potentials due to its smaller effective mass, $m^*_{InSb}/m^*_{InAs} = 0.57$. This effect is also seen in Fig. \ref{FIG8}(b), which shows larger subband separations than the InAs results shown in \ref{FIG5}(a). However, while the reduced subband occupation of InSb may provide some advantage, note that the gate voltage at which holes begin to emerge (dashed lines in Fig. \ref{FIG8}) is shifted to significantly less negative values compared to InAs due to the reduced bandgap of the InSb. This small bandgap inhibits InSb nanowires of reaching the few subband occupation regime except for small $\widetilde{V}_{SC}$, just as in the InAs case. Lastly, we note that non-parabolicity effects are stronger in InSb due to the small band gap. Therefore, the subband separation in the large subband occupation regime may be significantly reduced from the values predicted by the effective mass model used here. We conclude that, although the lower effective mass may provide some advantages, using InSb does not automatically solve the high-occupancy problem in Majorana devices. Since this is a potentially serious problem, it should be further investigated, both theoretically and experimentally, with the goal of identifying controlled procedures for driving the system in the few-subbands regime.   

\section{Local density of states} \label{LDOS}
To calculate the LDOS of states, we first note that the dispersion of the $n^{th}$ subband is given by
\begin{equation}
E_n\left(k\right) = E_{n,o} + \frac{\hbar^2 k^2}{2m^*},
\end{equation}
where $E_{n,o}$ is the energy of the subband bottom. This translates into a density of states of 
\begin{equation}
D_n(E) = \frac{1}{2\pi} \sqrt{\frac{m^*}{\hbar^2}} \frac{\Theta\left(E - E_{n,o}\right)}{\sqrt{E-E_{n,o}}} , \label{DOS}
\end{equation}
where $\Theta(E)$ is the Heaviside step function. We note the strong 1D square-root van Hove singularity in Eq. \ref{DOS}. The LDOS is then given by multiplying the density of states of each subband by the weight of the wavefunction the region of space of interest. The LDOS is given explicitly by 
\begin{equation}
LDOS\left(E\right) = \sum_n D_n\left(E\right) \int_\Omega \left|\psi_n (\vec{r})\right|^2 d\vec{r},
\end{equation}
where the sum runs over all subbands and $\Omega$ is the region of interest. We note that the wavefunction of each subband is independent of $k$ for this effective mass model. To eliminate the singularity in Eq.(\ref{DOS}) coming from the subband bottoms, we average the density of states over a small energy interval $\delta E = 0.2$ meV.

\bibliography{Draft3.bbl}

\end{document}